\documentclass[review,number,sort&compress]{elsarticle}

\usepackage{lineno,hyperref}
\usepackage{comment}
\usepackage{color}
\usepackage{siunitx}
\DeclareSIUnit[number-unit-product = ]\percent{\char`\%}
\usepackage{hyperref}
\usepackage[nowatermark]{fixmetodonotes}

\journal{Nuclear Instruments and Methods}

\begin{document}

\begin{frontmatter}

\title{Absolute Calibration of the DANCE Thermal Neutron Beam using Sodium Activation}

\author[Davis]{V. Fischer\corref{mycorrespondingauthor}}
\cortext[mycorrespondingauthor]{Corresponding author}
\address[Davis]{Department of Physics, University of California, Davis, CA 95616, USA}

\ead{vfischer@ucdavis.edu}

\author[Davis]{L. Pagani}

\author[Davis]{L. Pickard}

\author[Boston]{C. Grant}
\address[Boston]{Department of Physics, Boston University, Boston, MA 02215, USA}

\author[Davis]{J. He}

\author[Davis]{E. Pantic}

\author[Davis]{R. Svoboda}

\author[LANL]{J. Ullmann}
\address[LANL]{Los Alamos National Laboratory, LANSCE, Los Alamos, NM 87545, USA}

\author[Davis]{J. Wang}


\begin{abstract}
The measurement of the neutron capture cross section as a function of energy in the thermal range requires a precise knowledge of the absolute neutron flux. 
In this paper a method of calibrating a thermal neutron beam using the controlled activation of sodium is described. 
The method is applied to the FP-14 Time Of Flight neutron beam line at the Los Alamos Neutron Science Center to calibrate the beam to an uncertainty level of $\pm5$\,\%.
\end{abstract}


\end{frontmatter}



\section{Introduction}
\label{sec:intro}

The measurement of an isotope's neutron capture cross section relies on a precise understanding of the absolute neutron flux in the energy range of interest. 
In performing such a measurement, by thermal neutron activation in a reactor, the absolute neutron flux at thermal energies is obtained through the use of control samples of known composition, mass and cross section.
In the case of a Time Of Flight (TOF) neutron beam, such as Flight Path 14 (FP-14) at Los Alamos Neutron Science Center (LANSCE), neutron flux monitors are placed in the beam path and, through various neutron interactions depending on the monitor's neutron-sensitive isotope, provide the relative neutron rate as a function of neutron energy.
Although such monitors provide an accurate shape of the neutron flux energy profile from the beam TOF information, the absolute normalization of the neutron flux is dependent on many intrinsic parameters such as the geometrical acceptance and the detection efficiency.
Hence, these monitors must undergo periodic calibration in order to ensure they provide an accurate measurement of the absolute rate.
While calibrations at keV energies have been achieved by exposing samples with well-known resonance peaks~\cite{Albert:2016vmb}, calibration at thermal energies has not been done to a similar level of precision and accuracy. 

In this paper, a method to perform a thermal flux calibration using the controlled activation of a sodium sample, of known composition, mass, and geometry, is described. 
This study was performed at the FP-14 neutron beam at LANSCE.
However, the method can be reproduced at any TOF-based neutron beam facility.


\section{Measurement Strategy}
\label{sec:strategy}

A common method used to calibrate and measure the thermal neutron flux of a nuclear reactor is to perform neutron irradiation studies with control samples.
An example of this approach is Neutron Activation Analysis (NAA)~\cite{Greenberg:2011}.
Different isotopes, in different phases, can be used depending on the sample deployment setup, the thermal/fast neutron flux ratio, the isotope availability, and the desired activity to be achieved. 
Two other important criteria are the decay time and the precision to which the capture cross section is known both inside and outside the Region Of Interest (ROI).

In a typical neutron irradiation measurement, control samples are irradiated for periods ranging from minutes to several hours.
Their activity is subsequently measured on a high precision detector for hours to several days.
One must take into account these periods when choosing a suitable control sample - isotopes having half-lives on the order of tens of hours are usually preferred.
Another characteristic to consider is the neutron capture cross section.
Although the ambient neutron flux in a reactor and, for this study in LANSCE's FP-14, is strongly peaked at thermal energies, neutrons with energies ranging from the epithermal to the fast domains contribute to the total flux. 
In order to keep the contribution of these neutrons to the control sample activation down to negligible levels, it is sensible to choose a control isotope whose neutron capture cross section follows a pure ``1/v law''~\cite{Westcott:1960effective} with no capture resonances below intermediate neutron energies ($\sim 1$\,keV).
An element satisfying the aforementioned characteristics, and commonly used in nuclear reactors, is sodium:
\begin{itemize}
\item It has only a single stable isotope, $^{23}$Na.
\item It can be obtained in the inexpensive and easy to handle form of sodium carbonate, Na$_{2}$CO$_{3}$, with no long-lived isotopes from carbon capture and no hydrogen to scatter thermal neutrons.
\item Its daughter, $^{24}$Na, from the reaction $^{23}$Na+$^{1}$n$\rightarrow ^{24}$Na, has a $14.997$\,hour half-life, which is convenient for counting.
\item It has a resonance-free capture cross section below $500$\,eV. Therefore as a $1/v$-absorber, when subjected to a flux decreasing in this energy region, activation will predominantly be induced by thermal neutrons.
\end{itemize}

The basic measurement strategy was as follows: (1) fabricate a dry sodium carbonate target disk with a diameter larger than the neutron beam spot and a uniform, well-known thickness; (2) install the target in the FP-14 beam and irradiate for a known period, with the relative flux shape measured by the neutron beam monitors and the Protons on Target (POT) measured by the LANSCE beam current monitor; and (3) count the target to measure the amount of $^{24}$Na produced.

In analyzing the data, the $^{24}$Na production from three neutron energy regions are considered: (1) Region 1, the thermal region below $0.2$\,eV where the cross section is well known and the flux is high; (2) Region 2, the region from $0.2$ to $500$\,eV where there are not expected to be any resonances and the flux is known to be much lower but dominated by the non-thermal tail; (3) Region 3, the energy region above $500$\,eV where there are poorly know resonances, but where the flux is very low. 
As discussed in Sec.~\ref{sec:analysis}, $>87$\,\% of the captures are expected to come from Region 1, which will be crucial in determining the final uncertainly of this calibration technique.


\section{Experimental Setup}
\label{sec:setup}


\subsection{FP-14 Neutron Beam}
\label{sec:beam}

\begin{figure}[tb]
\centering 
\includegraphics[width=0.75\textwidth]{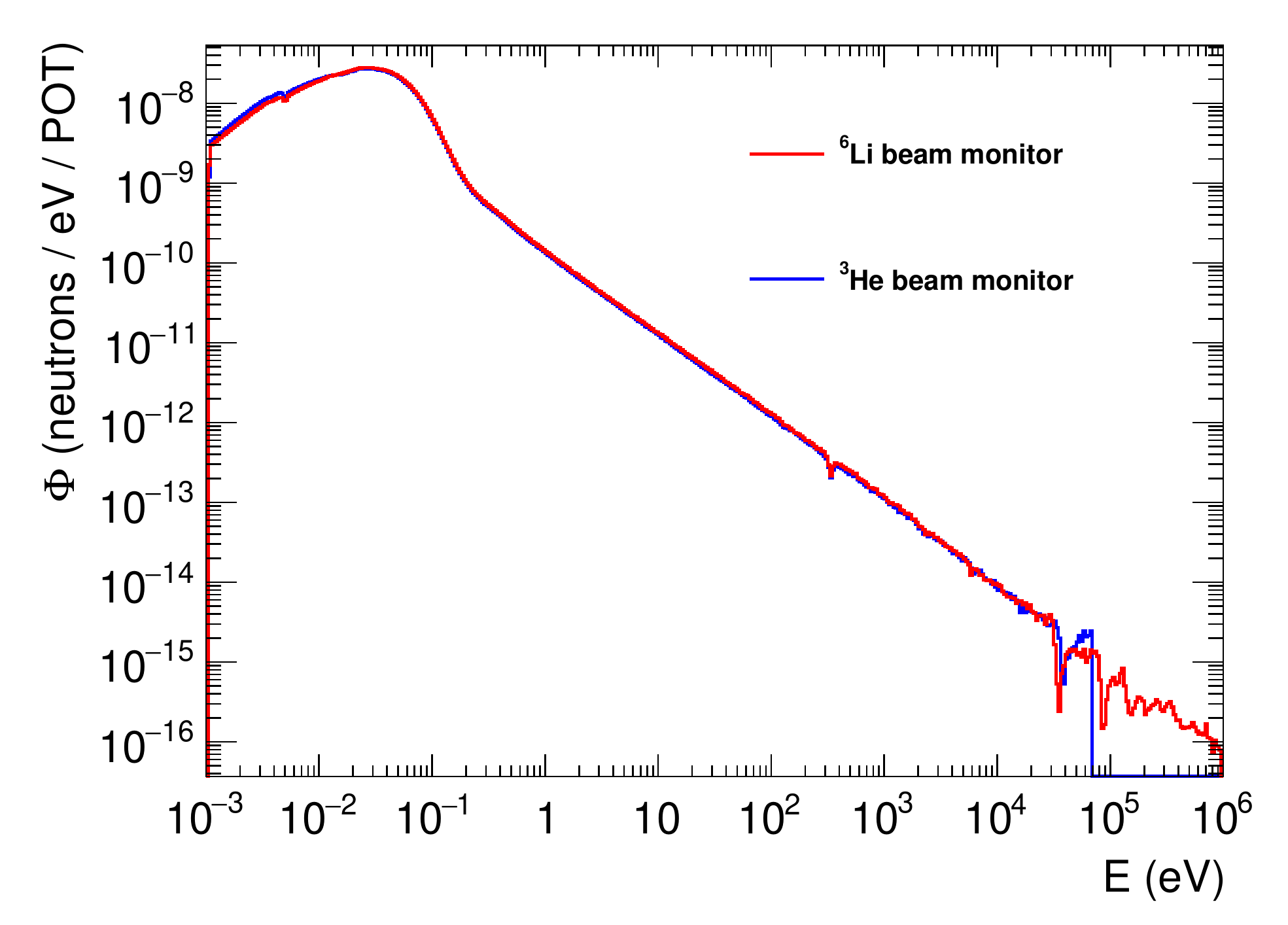}
\caption{Energy distributions of the LANSCE FP-14 neutrons as detected by the $^{6}$Li and $^{3}$He beam monitors. Both distributions are normalized using the method described in Sec.~\ref{sec:analysis}.}
\label{fig:beam_flux}
\end{figure}

The FP-14 neutron beam is located at the Manuel Lujan Jr. Neutron Scattering Center at the LANSCE~\cite{LANSCEBeam:1990}.
Neutrons originate from the spallation caused by a $800$\,MeV proton beam focused on a tungsten target. 
Typical proton currents in normal beam operation are approximately $100$\,$\mu$A.
The fast neutrons created are moderated in a water volume in such a way that the tungsten target is not in the direct line of sight of FP-14, thus lowering the probability of transporting un-moderated fast neutrons down to the experimental areas.
This setup allows neutrons traveling the beam pipe to have an energy spectrum strongly peaked at thermal energies.
The energy distribution of these neutrons is provided by two beam monitors, $^{6}$Li and $^{3}$He, through the observation of the $^{6}$Li(n,$\alpha$)$^{3}$H and $^{3}$He(n,p)$^{3}$H reaction rates, respectively.
The flux distributions as a function of the neutron energy for both monitors is shown on Fig.~\ref{fig:beam_flux}.
In the rest of the paper the $^{6}$Li monitor will be used as a reference, with the $^{3}$He monitor providing a cross-check.

One of the experimental areas of FP-14 hosts the Detector for Advanced Neutron Capture Experiments (DANCE)~\cite{Heil:2001dngcm}, located $20.25$\,m downstream from the water moderator.
DANCE is a $\sim$4$\pi$ gamma ray calorimeter made of a spherical array of 160\,BaF$_2$ crystals, each monitored by a photomultiplier tube. 
A detailed description of the detector can be found in Ref.~\cite{Reifarth:2013xny}.

Although the calibration method described in the following is independent of DANCE, due to geometrical considerations and easiness of access, the center of the detector was used as the irradiation location.


\subsection{Target Construction}
\label{sec:target}

\begin{figure}[tb]
\centering
\includegraphics[width=0.45\textwidth]{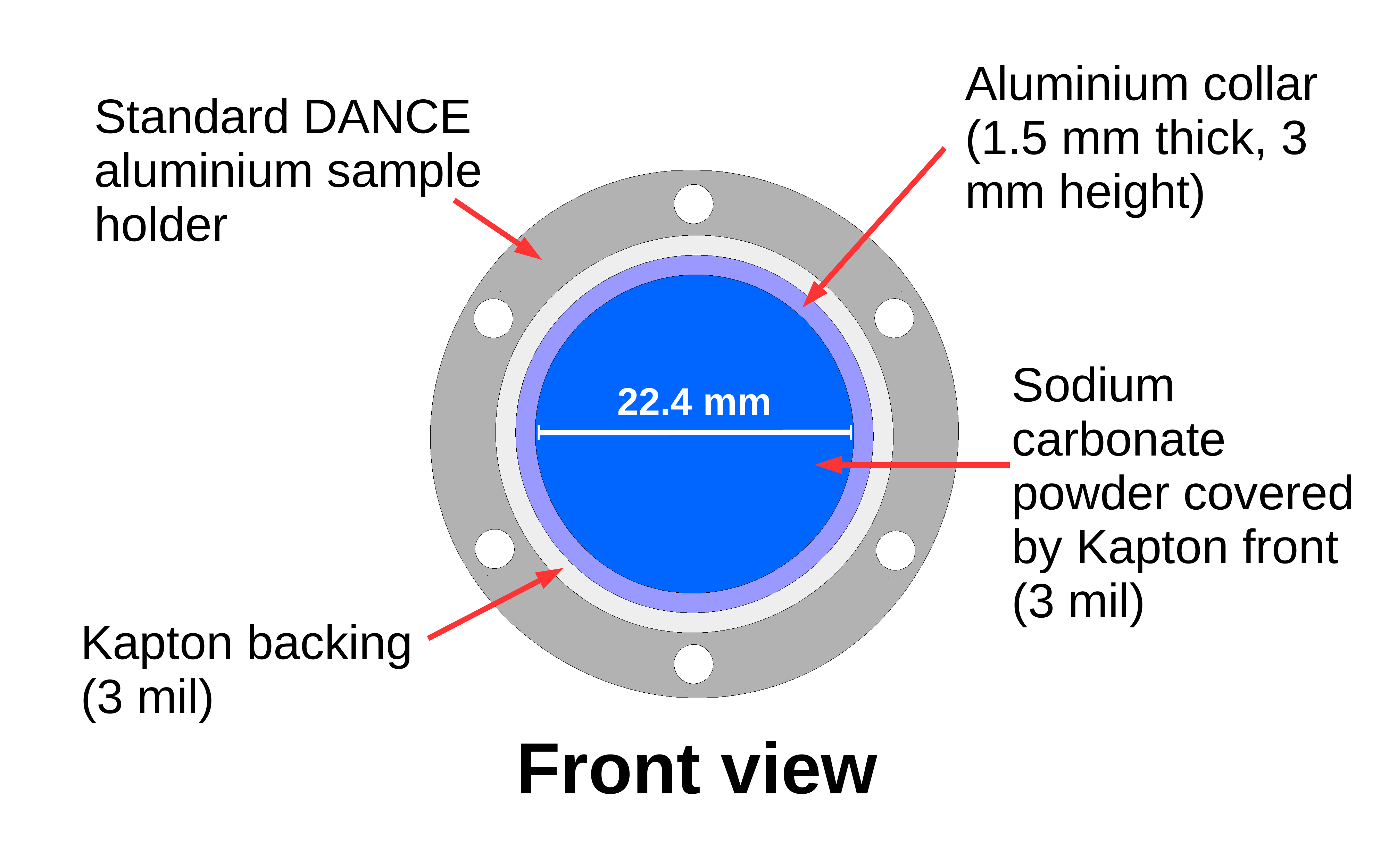}
\includegraphics[width=0.45\textwidth]{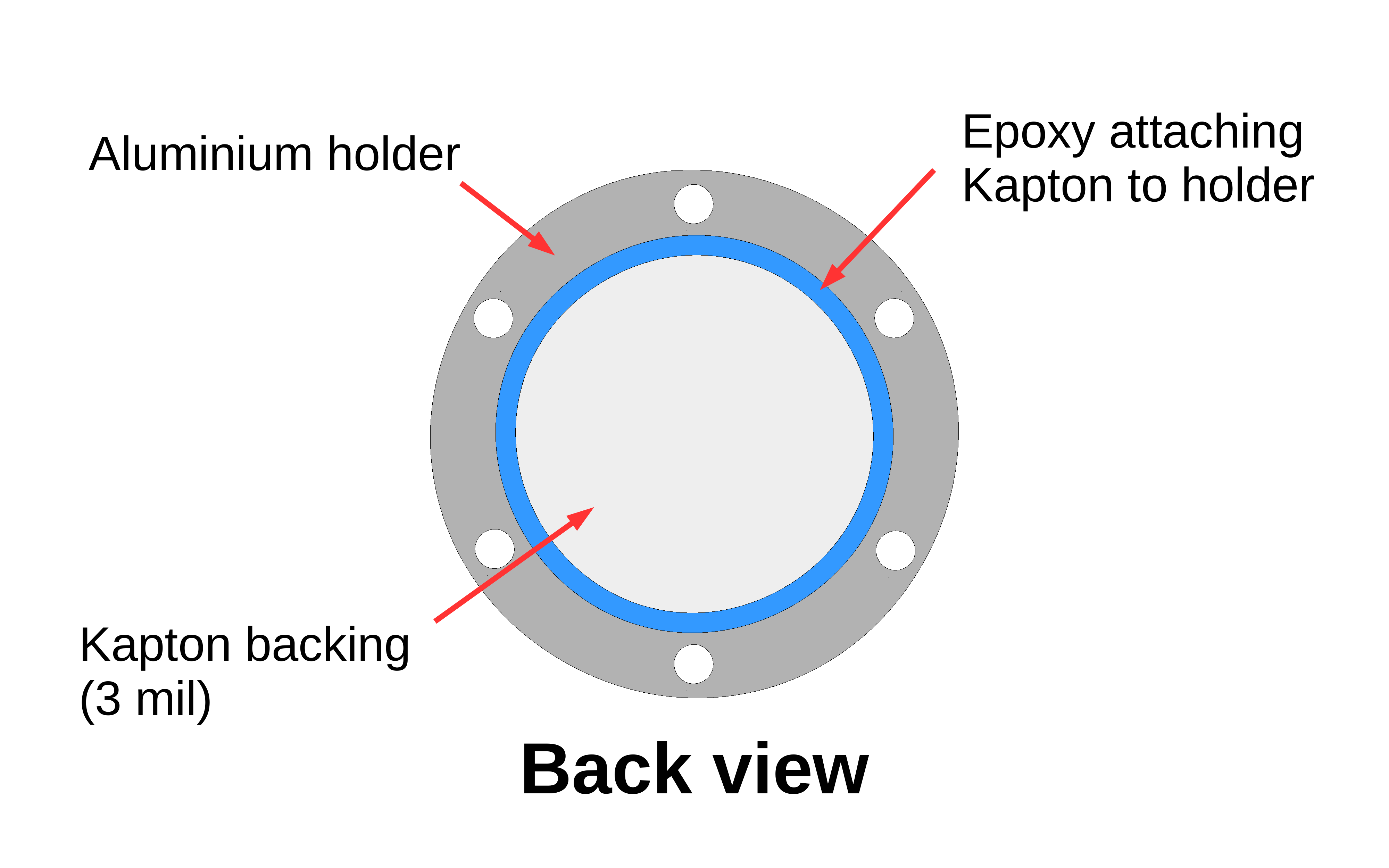}
\caption{The sodium target front (left) and back (right). The target was filled with anhydrous sodium carbonate that was loaded in a dry box to avoid water absorption.}
\label{fig:Target}
\end{figure}

The DANCE collaboration regularly deploys targets and radioactive sources at the center of the detector in order to perform beam profile and energy response calibrations.
In the former case, a target made of a Kapton film, onto which a thin layer of gold was deposited, secured with an aluminium ring is held in a dedicated holder.
This design, given its robustness and easiness of implementation, was chosen as the base design for the activation control assembly presented in this paper.
However, given the relatively low neutron capture cross section at thermal energies of sodium with respect to gold, a thicker target was necessary.

As shown in Fig.~\ref{fig:Target}, the sodium target was made of a $0.08$\,mm thick Kapton backing film onto which a $3$\,mm high aluminum ring had been attached.
With an inner diameter of $22.4$\,mm, the ring fully encompassed the $16$\,mm beam diameter~\cite{BeamProfileDANCE:2006} and the neutron flux outside the boundaries of the sodium target was considered negligible.
Upon being filled with the target material, a second Kapton window was epoxied on top of the aluminum ring in order to seal it.

So as to lower the probability of neutron scatters on hydrogen and maximize the density of sodium atoms in the target, anhydrous sodium carbonate, in a powder form, was selected as the target material.
Given its affinity for capturing water molecules, it had to be kept in a dry environment at all times during the target construction process.

The mass of target material was measured during the construction process.
From the known volume of the target and the molar masses of all atoms present in sodium carbonate, one can calculate the atomic density of $^{23}$Na in the target.
This density was determined to be $1.64 \times 10^{22}$\,atoms/cm$^{3}$.
A $1.6$\% correction was applied to this value in order to account for the fact the density of sodium bicarbonate powder in the target increased slightly when the target was placed vertically in the beam pipe due to gravity.
This effect led to the formation of a visible sub-mm high air gap in the target. 
The size of the air gap was accounted for in the correction and the uncertainty on its size propagated to the overall uncertainty on the $^{23}$Na atomic density, estimated at $2.96$\,\%.


\subsection{Gamma-ray Spectrometer}
\label{sec:HPGe}

The gamma ray spectrometer used to measure the activated sodium control upon irradiation was an ORTEC High Purity Germanium (HPGe) detector.

Prior to measuring the activity of the irradiated sodium sample, the HPGe counter was calibrated in terms of energy response and detection efficiency.
The latter was performed using a radioactive source placed 25~cm away from the face of the HPGe crystal enclosure.
For geometrical consistency, the sodium target was placed at the same position during the activity measurement.

In theory, since the only gamma line of interest during the target counting is the 1368.6~keV ray from $^{24}$Na, one needs only to obtain the HPGe detection efficiency at this energy.
However, without access to a calibrated $^{24}$Na source, the sensible way to obtain the detection efficiency at $1368.6$\,keV is to interpolate the efficiency curve obtained using several other data points spread over a wide range of energies.
For that purpose, a ``mixed" source containing nine different radioactive isotopes in various concentrations was used.
Due to the age of the source and the low initial activity of some isotopes, only four isotopes were considered in the efficiency calibration.
The isotopes of interest along with their respective gamma lines were $^{113}$Sn ($391.7$\,keV), $^{137}$Cs ($661.7$\,keV), $^{60}$Co ($1173.2$ and $1332.5$\,keV) and $^{88}$Y ($898.0$ and $1836.1$\,keV).
The use of a $^{60}$Co source and its $1332.5$\,keV gamma line is of particular interest given the proximity to the main sodium gamma line.
The activities of all isotopes present in the source have been measured by NIST with a $3.1$\,\% uncertainty. 
In the following, these uncertainties were conservatively considered to be correlated by assuming the same instrument was used to measure the activity of all sources.

The HPGe efficiency to detect the main sodium gamma ray was obtained from the evaluation of the second-degree polynomial function fitting the data points at $1368.6$\,keV.
This value which combines both geometrical and intrinsic detection efficiencies, was found to be equal to $0.033$\,\% with an uncertainty of $3.1$\,\%, entirely dominated by the uncertainty on the activities of the calibrations sources.

Given the geometrical difference between the radioactive source and the sodium target, one has to quantify the difference in relative solid angle, with respect to the HPGe, between the two.
Using Monte Carlo, the discrepancy between the two solid angles was found to be $0.26$\,\% and was taken into account in the calculation of the neutron flux systematic uncertainties.
The discrepancy between the transmission of gamma rays in the source and in the target was found to be negligible as both are $99$\,\%.


\section{Data Analysis and Results}
\label{sec:analysis}

The sodium target was placed at the center of the DANCE detector array and irradiated for $11$\,hours and $7$\,minutes. 
During this period, the neutron beam was down for $24$\,minutes thus yielding an effective irradiation time of $38,580$\,seconds ($t_i$).
This effect of this downtime has been taken into account in the following calculation.
The number of beam spills was recorded and provided a second independent measurement of the irradiation time after correction with the known beam frequency of $20$\,Hz.
The difference between these two measurements was taken as the uncertainty on the irradiation time and amounts to $0.6$\,\%.
After irradiation, the sodium target was transferred to the HPGe counting facility.
This transfer, whose duration must be accounted for given the relatively short half-life of $^{24}$Na, took $2,580$\,seconds ($t_t$) with negligible uncertainty.
The sodium target activity was measured using the HPGe detector for a total time of $45,688$\,seconds ($t_c$), with a dead time of $0.9$\,\%, giving a live time of $45,599$\,seconds, with negligible uncertainty.

The shape of the neutron energy distribution was extracted from the beam monitors and standard counting techniques were then used to calculate the number of captures up to a normalization constant. 
The total number of gamma events ($G$) detected in the ROI is related to the $^{23}$Na neutron capture cross section ($\sigma$), and to the neutron flux ($\phi$), as:  
\begin{equation}
\label{eq:naa}
\sum_i \phi_i \, \sigma_i = \frac{G}{\varepsilon_{\gamma} \, I_{\gamma}} \, \frac{\lambda e^{\lambda t_t}}{\left(1-e^{-\lambda t_c}\right) \, \left(1-e^{-\lambda t_i}\right)} \, \frac{1}{\rho \, l}
\end{equation}
where the sum is over all \textit{i}-energy bins measured by the beam monitors. 

The variables $\varepsilon_{\gamma}$, $I_{\gamma}$, and $\lambda$ are the detection efficiency of the HPGe in the ROI (see Sec.~\ref{sec:HPGe}), the branching ratio for the $1368.6$\,keV line ($99.99$\,\%), and the decay constant of $^{24}$Na ($1.284 \times 10^{-4}$ \,s$^{-1}$), respectively. 
$\rho$ and $l$ are the density of $^{23}$Na atoms (($1.64 \pm 0.05) \times 10^{22}$\,atoms/cm$^{3}$) and the target thickness ($3$\,mm), respectively.
The uncertainties on $I_{\gamma}$, $\lambda$, and $l$  are considered to be negligible.
The exponential terms account for the radioactive isotopes, created upon irradiation, constantly decaying throughout the irradiation, transfer and counting periods.

Coincident detection of $2754.0$\,keV (with a branching ratio of $99.86$\,\%) and $1368.6$\,keV lines from $^{24}$Na, cause the latter to be unaccounted for in Eq.~\ref{eq:naa}. By calculating the relative solid angle between the target and the HPGe detector one can estimate this effect.
Conservatively not including detection efficiency, it is found to be $0.36$\,\% and is taken into account in the calculation of the neutron flux systematic uncertainties.

\begin{figure}[tb]
\centering 
\includegraphics[width=0.75\textwidth]{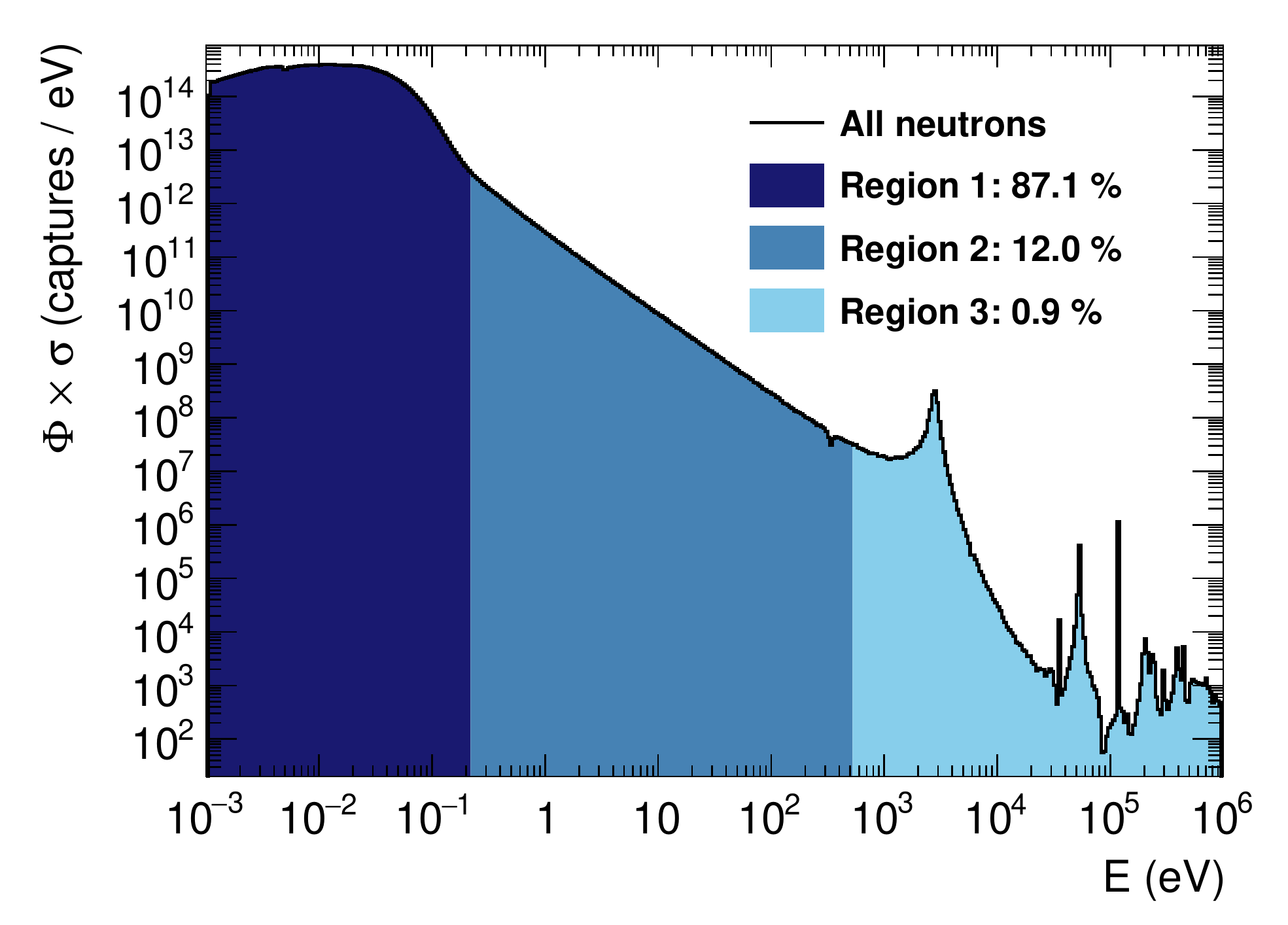}
\caption{Product of the FP-14 neutron flux and the $^{23}$Na capture cross section normalized using the result from Eq.~\ref{eq:naa}. The three domains highlighted on the spectrum corresponds to the three energy regions described in Sec.~\ref{sec:strategy}.}
\label{fig:beam_norm_3regions}
\end{figure}

Fig.~\ref{fig:beam_norm_3regions} shows the product of the $^{23}$Na neutron capture cross section and the neutron flux once normalized using the result from Eq.~\ref{eq:naa}.
The thermal region, Region~1, below $0.2$\,eV, provides $87.1$\,\% of the neutrons involved in the activation of the $^{23}$Na.
Region~2 provides $12.0$\,\% of the neutrons contributed to the $^{24}$Na production, while Region~3, the resonance region, only contributes to the activation at the level of $0.9$\,\%. Thus this method is especially effective in calibrating the thermal neutron flux, where the uncertainty in the capture cross section is small. For example, at $25.3$\,meV, it is known to be $530 \pm 5$\,mb - where the systematic uncertainty accounts for the discrepancy between previous measurements~\cite{BROWN20181}.

In Region 1, the small discrepancy in shape between the $^{6}$Li and the $^{3}$He beam monitors results in a $2.1$\,\% systematic uncertainty on the total number of thermal neutrons (see Fig.~\ref{fig:beam_flux}).

As the beam monitors are not sensitive to neutrons with energies below $1$\,meV, the effect of this limit in range on the calculated total number of thermal neutron was evaluated by extrapolating the measurements of the beam monitors to lower energies assuming a ``$v \, f_{MB}(v)$'' response (where $f_{MB}(v)$ is the Maxwell-Boltzmann distribution and $v$ is the velocity of the neutron).
The extrapolation is found to be in good agreement ($<1$\,\% discrepancy) with the beam flux obtained from Ref.~\cite{LowE_beam_extrapol}. 
This is a small effect, as these neutrons are expected to account for $0.73$\,\% of the expected sodium activation, but it is nevertheless included. 

The final result of this analysis yields $(5.04 \pm 0.08 \text{ (stat.)} \pm 0.25 \text{ (sys.)}) \times 10^{10} $ thermal neutrons in Region~1 for the total irradiation period, which consisted of 776,230 beam pulses.
Tab.~\ref{tab:xsec_sys_errors} summarizes the contributions to the statistical and systematic uncertainties included in this result.

\begin{table}[tb]
\caption{Summary of the contributions to the neutron flux uncertainty.}
\begin{center}
\begin{tabular}{ccc}
Error & Stat. (\%) & Sys. (\%) \\ 
\hline
HPGe efficiency ($\varepsilon_{\gamma}$) & negligible & 3.1 \\
Target density ($\rho$) & 0.0 & 3.0 \\
Beam monitors shape discrepancy & 0.0 & 2.1 \\
Gamma ray signals ($G$) & 2.0 & 0.0 \\
Capture cross section ($\sigma_{25.3 meV}$) & 0.0 & 0.9 \\
Low energy neutron contribution & 0.0 & 0.7 \\
Irradiation time ($t_i$) & 0.0 & 0.6 \\
Summing effects from the $2754.0$\,keV line & 0.0 & 0.4 \\
Source-target solid angle discrepancy & 0.0 & 0.3 \\
\hline
\end{tabular}
\end{center}
\label{tab:xsec_sys_errors}
\end{table}

Since the intensity of the beam at FP-14 fluctuates over time, the recorded number of POT can be used in order to relate the results obtained during this sodium calibration to any other irradiation.
This is made possible by the fact the number of neutrons, and consequently the number of thermal neutrons, is proportional to the number of POT.
During the sodium target irradiation, the LANSCE beam facility recorded a total of $2.44 \times 10^{19}$\,POT yielding a thermal neutron rate of $(2.07 \pm 0.03 \text{ (stat.)} \pm 0.10 \text{ (sys.)}) \times 10^{-9}$\,neutrons per POT. 
Thus this result can be used to calibrate future experiments in this beam line.


\section{Summary}
\label{sec:summary}

Using a sodium control sample, the absolute thermal neutron flux of FP-14 at LANSCE has been measured to be $2.07 \times 10^{-9}$\,neutrons per POT with an associated uncertainty of $5$\,\%.
Although this thermal flux is specific to the FP-14 beam, assuming similar beam conditions, the method described in this study is interchangeable and can be used to normalize any other thermal neutron beam.

This work was supported by the U.S. Department of Energy (DOE) Office of Science under award number DE-SC0009999, and by the DOE National Nuclear Security Administration through the Nuclear Science and Security Consortium under award number DE-NA0003180. 
We gratefully acknowledge the logistical and technical support and the access to laboratory infrastructure provided to us by LANSCE and its personnel at the Los Alamos National Laboratory. 


\section{Bibliography}

\bibliographystyle{unsrt}
\bibliography{bibliography.bib}

\end{document}